\documentstyle{article}

\def\QQa{\renewcommand{\baselinestretch}{1.3}\Huge\normalsize\large\small}

\begin{document}
\QQa

\large
\begin{flushright}
ITP.SB-92-62\\
Nov.9,1992
\end{flushright}

\begin{center}
\huge
The Canonical Quantization in Terms of Quantum Group and
Yang-Baxter Equation\\
\vspace{1cm}
\Large

Chang-Pu Sun \footnote {\large
Permenet address:Physics Dpartment,Northeast Normal University,
Changchun 130024,P.R.China}\\

Institute for Theoretical Physics,State University of New York,Stony Brook,
NY 11794-3840,USA\\

\vspace{1cm}
\huge
Abstract\\
\end{center}

\large
In this paper it is shown that a quantum observable algebra, the
Heisenberg-Weyl
algebra, is just given as the Hopf algebraic dual to the classical
observable algebra over  classical phase space and the Plank constant is
included in this
scheme of quantization as a compatible parameter living in the quantum double
theory.In this sense,the quantum Yang-Baxter equation naturally appears as
a necessary condition to be satisfied by a canonical elements,the universal
R-matrix,intertweening the quantum  and classical observable algebras.
As a byproduct,a new ``quantum group'' is obtained as the quantum double of the
classical observable algebra.

\newpage
\large

The purpose of this paper is trying to understand  directly
physical meaning of the
quantum group theory [1] in the basic quantum mechanics.We formally reproduce
the canonical quantization from the quantum double of the commutative classical
 observable algebra(COA) A for the  classical phase space.It has to be pointed
out that the quantum group theory originally comes from the quantization of
some
non-linaer problems in physics(such as the S-matrics theory in low dimensional
quantum field thoery[2] and quantum inverse scattering methods[3]) and
exactly-solvable
models in statistical mechanics [4].

Let us consider a classical observable algebra associated with the classical
phase space with two canonical variables,the coordinate Q and the momentum
P.The COA A is an associative algebra generated by P,Q and a central elements N
.In the classical sense,P and Q commute each other,so the COA is an Abelian
algebra and
possesses simplest structure as an associative algebra.Fortunately,this
 simplest algebra can be endowed with a `quantum group' structure $(\Delta,
S ,\epsilon)$
$$\Delta(x)=x\otimes 1 + 1\otimes x,~~x=P,Q$$
$$\Delta(N)=N\otimes 1+1\otimes N+\mu P\otimes Q ,\eqno{(1)} $$
$$S(x)=-x,S(N)=-N+\mu PQ,S(1)=1,x= P,Q$$
$$\epsilon(y)=0,y=P,Q,N,\epsilon (1)=1$$
where $\Delta,\epsilon$ are algebraic homomorphism and S an algebraic
antihomomorphism ;$\mu$ is a compatible complex parameter for  $(\Delta,
S ,\epsilon)$ satisfying the axioms of Hopf algebra .With the above
 structure $(\Delta,S ,\epsilon)$ the COA becomes
a commutative, but non-cocommutative Hopf algebra if $\mu$  is not zero.

Now,we set to find the quantum daul(also called Hopf( algebraic) dual [1]) B
of the COA A according to Drinfeld's quantum double  theory(for the reviews
easy to physicists please see the refs.[5,6]).To this end,we define the
generators $\hat{ P},\hat{Q}$ and E by
$$<P^mQ^nN^l,\hat{P}>=\delta_{m,1}\delta_{n,0}\delta_{l,0 }$$
$$<P^mQ^nN^l,\hat{Q}>=\delta_{m,0}\delta_{n,1}\delta_{l,0 },\eqno{(2)}$$
$$<P^mQ^nN^l,E>=\delta_{m,0}\delta_{n,0}\delta_{l,1 }$$
where the bilinear form $<,>:A\times B\rightarrow complex~ field~ C$
sasisfies the following conditions resulting  from the duality in quantum
double:

$$<a, b_1 b_2>=<\Delta_A (a), b_1\otimes b_2>, a\in A,~b_1,~b_2\in B,$$

   $$<a_1 a_2,b>=<a_2\otimes a_1,\Delta_B(b)>,a_1,a_2\in A,b\in A $$

$$<1_A,b>=\epsilon_B(b),b\in B,\eqno{(3)}$$
  $$<a,1_B>=\epsilon_A(a),a\in A$$
$$<S_A(a),S_B(b)>=<a,b>,a\in A,b\in B$$
In this letter,without confusion,we do not specify the operation
($\Delta,S,\epsilon$) for A and B.

It follows from eqs.(2) and (3) that
$$<P^mQ^nN^s,\hat{P}^k\hat{Q}^lE^r>=m!n!s!\delta_{m,k}\delta_{n,l}\delta_{s,r
}$$
$$\Delta(x)=x\otimes 1+1\otimes x,S(x)=-x,\epsilon(x)=0\eqno{(4)}$$
$$\epsilon(1)=1,S(1)=1,x=\hat{P},\hat{Q},E$$
The key to our study is the commutation relations beween $ \hat{P}~and~
\hat{Q}$ .From eqs(1),(2) and (3),we derive
$$<P^mQ^nN^l,\hat{P}\hat{Q}>=<\Delta(P^mQ^nN^l),\hat{P}\otimes \hat{Q}>$$
$$\sum^{m}_{k=1}\sum^{n}_{r=1}\sum^{l}_{s=1}\sum^{l-s}_{t}\frac{m!l!n!\mu^s}
{k!r!s!(m-k)!(n-r)!(l-s-t)!t!}$$
$$<P^{m-k+s}Q^{n-r}N^{l-s-t}\otimes P^kQ^{r+s}N^t,\hat{P}\otimes \hat{Q}>
=\mu^{\delta_{l,1}}\delta_{m+l,1}\delta_{n+l,1}$$
$$<P^mQ^nN^l,\hat{Q}\hat{P}>=\delta_{m,1}\delta_{n,1}$$
that is
$$<P^mQ^nN^l,[\hat{P},\hat{Q]}>=\mu\delta_{l,1}\delta_{m,0}\delta_{n,0}$$
Then,we obtain
$$[\hat{P},\hat{Q}]=\mu E,$$
$$[\hat{P},E]=0=[\hat{Q},E],\eqno{(5)}$$

Then,we show that {\bf
 the Hopf duals $\hat{P}~ and~ \hat{Q}$ of the classical
canonical coordinate Q and momentum P are just the quantum coordinate and
momentum operators respectively if we can take $\mu=i\hbar$.Therefore,
 the Drinfeld's quantum double theory
provides us with an algebraic scheme of the canonical quantization in the
basic quantum mechanics!} .In fact, the parameter $\mu$ characterizes
the degrees of the non-cocommutation of the classical observable algebra A
 and the non-commutation of the quantum observable algebra(QOA) B generated by
$\hat{P},\hat{Q}$ and E(it is usaully called Heisenberg-Weyl(HW) algebra)
.Since when
$\mu$ is zero,the algebra B becoms commutative,it is reasonable to take
$\mu=i\hbar$.In this way,the Plank constant $\hbar$ automatically
,enters the algebra B
to realize the `algebraic' canonical quantization.Notice that with the
the Hopf algebraic strucure (4),the QOA B is a cocommutative,but non-
commutative Hopf algebra.

In order to investigate the quantum Yang-Baxter equation
in connection with canonical quantization we need to combine A with B
to get a  qausi-trianglur Hopf algebra as the quantum
double of the COA A.Thanks to the double muiltiplication
rule given in Drinfeld's theory

$$ba=\sum_{i,j} <a_i(1),S(b_j(1))><a_i(3),b_j(3)>a_i(2)b_j(2) \eqno{(6)}$$
$$(id\otimes \Delta)\Delta (c)=\sum_i c_i(1)\otimes c_i(2)\otimes c_i(3)$$
we get the commutation relations between A and B

$$[E,everything]=0,$$
$$[N,\hat{P}]=-\mu Q,[N,\hat{Q}]=\mu P.\eqno{(7)}$$
Then,we obtain the universal R-matrix
$$ \hat{R}=\sum a_i\otimes b_i
=\sum^{\infty}_{m,n,s}\frac{P^mQ^nH^s\otimes \hat{P}^m\hat{Q}^nE^s}
{m!n!s!}\eqno{(8)}$$
$$=exp[ P \otimes \hat{P}]=exp[Q  \otimes \hat{Q}]exp[ N \otimes E ]
$$
as a canonical elements intertweening the QOA B and COA A.According to
Drinfeld's theory,the universal R-matrix constructed above must satisfies
the abstract Yang-Baxter equation

$$\hat{R}_{12}\hat{R}_{13}\hat{R}_{23}=\hat{R}_{23}\hat{R}_{13}\hat{R}_{12},
\eqno{(9)}$$
where $a_m$ and $b_m$ are the basis elements of A and B respectively,and they
are dual each other;$<a_m,b_n>=\delta_{m,n}$;\\
$$\hat{R}_{12}=\sum_m a_m\otimes b_m\otimes 1,\\
\hat{R}_{13}=\sum_m a_m\otimes 1 \otimes b_m,\\
\hat{R}_{23}=\sum_m 1\otimes a_m\otimes b_m
$$
Then,we can conclude that the classical observable algebra A can incorporates
the quntum  observable algebra B to form a new quantum group D
-a quasi-triangular
Hopf algebra ,which is generated by P,Q,N,E,$\hat{P}~and~\hat{Q}$ with the
relations  (1),(4) (5) and (7);the quantum Yang-Baxter equation as a necessary
condition is enjoyed in this construction of the canonical quatization.
In fact,the new quantum group D ,as an associative algebra,is the universal
enveloping algebra of the non-simple Lie algebra with the basis
  P,Q,N,E,$\hat{P}~and~\hat{Q}$.This means that ,besides so-called
q-deformations of some `classical algebra',one not only endows  some finite
dimensional
Lie algebras (strictly, their universal enveloping algebras)with a
cocommutative, but also with a non-cocommutative Hopf algebra strucure .The
discussion of this paper is a typical example of  exotic quantum
doubles of `non-q-deformation' and another type of such exotic quantum
double has been constructed in connection with infinite dimensional
Lie algebra [7].As for the quantum theory,the present studies shows
that the Drinfeld's quantum double theory can
provides us with an algebracrized scheme of quantization
available for both the basic quantum mechanics and the nonlinear
quantum systems such as in quantum inverse scattering method.

A direct generaliztion of present study is the higher-dimensional case where
we only need write down the non-cocommutative coproduct
$$\Delta(N)=N\otimes 1+1\otimes N +i\hbar \sum^{M}_{k=1}P_k\otimes
Q_k\eqno{(9)} $$
for N and the similar for other generators $P_k,Q_k$ (k=1,2,...,M) in the
muilti-COA A(M).Its quantum daul is just the M-state HW algebra with the
momentum and coordinate operators $\hat{P}_k,\hat{Q}_k$ and the unit element
E as generators,and there naturally is a new quantum double(`group')
as a routine multi-generalization of the algebra D obtained above.

Before concluding the discussion in this paper,we would like to understand
the physical meaning of the universal R-matrix (8) and the operator N.As for
the later,we can give N a realization in terms of   $\hat{P}~ and~ \hat{Q}$
$$N=P\hat{P}+Q\hat{Q}$$
which preseves all the commutators of N with other generators and where we have
take a representation with E to be unit.So it maybe
implies a `interaction' between the classical and quantum objects.For the
former,we need consider the representation theory of the quantum group D.
Thanks to the Schur lemma,the central elements P,Q and E must be some
scalars in an irreducible representation.In this sense,$P\otimes \hat{P},
 ~and~Q\otimes \hat{Q}$ act as $P \hat{P}~and~Q \hat{Q}$ on second component
of the product space $V\otimes V$ where V is the representation space.Thus,
the universal R-matrix has an equivelent form
$$\hat{R}=exp[P\hat{P}+Q\hat{Q}]\otimes exp[P\hat{P}]exp[Q\hat{Q}]\eqno{(10)}$$
where we have renormalized the scalar E to be 1.Using the Campbell-Hausdorff
formula,
we prove that the universal R-matrix is just the generating operator of
the two-model coherente state
$$\mid Z>=e^{\frac{i\hbar}{2} PQ}exp[Za^++Z^*a]\otimes exp[Za^++Z^*a]\mid
0>,Z=Q+iP
$$
that has not been normalized.This shows the possible
relatons beteen the coherent state
and the quantum Yang-Baxter equation.We also notice that in some
representations similar to what we considered above,the R-matrices
are completly factorized as the solutions for Yang-Baxter equation.

Finally,we point out that a class of new R-matrices ,as the
matrix representations of the universal R-matrix for the new quantum group
D,can follow from the finte
dimensional representations.To get them,we need the detailed discussions
for the general representation theory of D.These corresponding studies
with some mathematical interests will be published elsewhere.

\vspace{1cm}

\large
The author would like  to express his sincere thanks to
Prof.C.N.Yang for drawing his
attentions to the research field related to
the quantum Yang-Baxter equantion and for giving him very kindly helps.
He is supported by Cha Chi Ming fellowship through the CEEC in State
University of New York at Stony Brook,and also is supported in part by the
NFS of China through Northeast Normal University .He also thanks Prof.
Takhtajian and Dr.F.Yu for useful discussion.

\newpage

\huge
References\\

\large

\noindent
1.V.G.Drinfeld,Proc.ICM.Berkeley,1986,(ed.By A.Gleason,AMS,1987),p.798\\
M.Jimbo,Lett.Math.Phys.10(1985)63\\
L.D.Faddeev,N.Y.Reshetikhin,L.A.Takhtajian,Preprint LOMI E-l4-87,1987;also in
Algebraic analysis,vol.1(198801299\\
for a review see {\it Braid Groups,Knot Theory and Statistical
Mechanics} C.N.Yang,M.L.Ge(eds.),
Singapore:World Scientific .1989.\\
2.C.N.Yang,Phys.Rev.Lett.19(1967)1312\\
A.B.Zamolodchikov,Al.B.Zamolodchikov ,Ann.Phys.120(1979)253\\
H.J.de Vega,Inter.J.Mod.Phys.A 4(1989),2371.\\
3.E.K.Sklyanin,L.A.Takhtajian and L.D.Faddeev,Theor.Mathem.Fisica 40(1979)194\\
P.P.Kulish and N.Y.Reshetikhin,J.Phys.A.16(1983),L591\\
4.R.J.Baxter,{\it Exactly-Solved Models in Statistical Machanics}
Academic.Press,1982.\\
5.L.A.Takhtajian,{\it Quantum Groups}, in Nankai Lectures,1989,ed.by M.L.Ge
and B.H.Zhao,World Scientific,1990.\\
6.M.Jimbo,{\it Topics from representations of $U_q(g)$ }, in Nankai Lectures,
1991,ed.by M.L.Ge,World Scientific,1992.\\
7.C.P.Sun,X.F.Liu and M.L.Ge,J.Math.Phys.33(1992),in press.\\
 W.Li,C.P.Sun and M.L.Ge,Preprints ITP-SB.92-59,1992\\

\end{document}